# New sources of economies and diseconomies of scale in on-demand ridepooling systems and comparison with public transport

Andrés Fielbaum*[1] Alejandro Tirachini[2,3] Javier Alonso-Mora[1]
[1]**Department of Cognitive Robotics, TU Delft**
[2]**Civil Engineering Department, Universidad de Chile**
[3] Instituto Sistemas Complejos de Ingeniería, Chile
*Corresponding author, a.s.fielbaumschnitzler@tudelft.nl

### Abstract

On-demand ridepooling (ODRP) can become a powerful alternative to reduce congestion and emissions, if it attracts private car users. Therefore, it is crucial to identify the strategic phenomena that determine when ODRP systems can run efficiently. In this paper, we analyze the performance of an ODRP system, in which the fleet of low-capacity vehicles is endogenously adapted to the demand, and operated in a zone covered by a single transit line. The routing of the on-demand fleet follows some of the rules of public transport systems; namely, it is not-for-profit, some users can be required to walk, and all requests must be served. Considering both users' and operators' costs we identify two sources of scale economies: when demand grows, the average cost is reduced due to a) an equivalent of the Mohring Effect (also present in public transport), and b) due to matching users with more similar routes when they are assigned to the vehicles, which we call *Better-matching Effect*. A counter-balance force, called *Flex-route Effect,* is observed when the vehicle loads increase and users face longer detours. We find a specific demand range in which the latter effect dominates the others, imposing diseconomies of scale when only users' costs are considered. Such a phenomenon emerges because the routes are not fixed; hence, it is not observed in traditional public transport systems. However, when considering both users' and operators' costs, scale economies prevail. Our simulations show that relaxing door-to-door vehicle requirements to allow short walks is crucial for the performance of ODRP. In fact, we observe that an ODRP system with human-driven vehicles and walks allowed has a total cost at a similar level to that of a door-to-door ODRP system with driverless vehicles.

**Keywords:** On-demand mobility, Ridepooling, Scale economies, Public transport, Mohring effect, Flex-route effect, Better-matching effect

## 1. Introduction

### 1.1 On-demand ridepooling systems: Potential and challenges

Transport systems are facing profound transformations worldwide thanks to the ability to connect vehicles and large numbers of passengers on-demand. After some five years of their arrival, several studies have shown that transportation network companies (TNCs) have increased traffic and congestion without reducing vehicle ownership (Diao et al., 2021; Henao & Marshall, 2019; Roy et al., 2020; Tirachini & Gomez-Lobo, 2020; Ward et al., 2021; Wu & MacKenzie, 2021). This situation has fostered the study and implementation of on-demand *ridepooling* (ODRP) services, in which different users simultaneously share a vehicle when their routes are compatible, so that congestion and emissions can be reduced (W. Li et al., 2021; Tikoudis et al., 2021).

ODRP systems have the potential to lower congestion because they might reduce the required fleet significantly when compared to the non-pooled versions, as shown by several previous studies (Alonso-Mora et al., 2017; Fagnant & Kockelman, 2018; Santi et al., 2014). However, the respective analyses are based on comparing the number of vehicles needed to serve a fixed demand, which might be troublesome as it might not be that both systems attract the same users. In fact, recent papers suggest that the ability of ODRP to reduce congestion depends on reaching some advantageous scenarios (Ke et al., 2020; Tirachini et al., 2020). Such scenarios should combine an efficient fleet operation with an ability to attract passengers from private modes rather than from public transport. To reach those scenarios, some strategic decisions arise, such as whether it is efficient to use ODRP to replace the public transport network (an idea that has recently begun to be studied by scholars, as we explain in Section 2.3 when revising related works), or if it is better to use them in low-demand or high-demand areas.

These strategic questions require a deeper understanding of the operation and the virtues of ODRP systems. However, this is not an easy task, as the operation of ODRP depends on specific algorithms to face the complexity of operating on-demand and with a large number of feasible ways to match users and vehicles. Which algorithm to utilize may yield different strategic results and affect scale effects. For instance, a seminal study by Li & Quadrifoglio (2010) studies a last-mile service that dispatches vehicles sequentially as soon as they get enough users regardless of their destinations. When doing so, a potential source of scale economies is not leveraged, namely that a greater demand enables grouping together users with closer destinations without increasing waiting times significantly.

## 1.2 Overview, contributions, and structure of the paper

In this paper, we extend a state-of-the-art assignment model to perform a detailed economic analysis that uncovers the multiple dimensions that determine the efficiency of ODRP as a shared-mobility platform for urban operations, compared with a traditional public transport system.

To do this, we use an extended version of the so-called "single-line model", in order to study sources of economies and diseconomies of scale when operating ODRP. The traditional single-line model, which analyzes one transit line as isolated from the rest of the system, has been extensively used by researchers across decades to analyze structural aspects of public transport design. Its usefulness resides in being simple, as it permits studying the impact of the demand conditions (or other parameters) over the mobility system under scrutiny, excluding the spatial distribution. By this means, the demand can be represented by a single variable (or a few of them), which makes this model quite precise for scale analysis.

The single-line model is useful for scale analysis but has a relevant limitation when studying on-demand systems: vehicles' routes are not defined a priori but adapted to the emerging users. Such a feature cannot be captured in a single-line model in which there is only one possible route. This limitation might influence scale analysis, as one aspect to study is the evolution of the routes with scale (in fact, Manik & Molkenthin, 2020, show that a lineal network artificially favors the performance of ODRP over several alternative topologies). We address this limitation by extending the single-line model, so that we keep most of its simplifying aspects, but yet enabling different routes to be followed depending on the passengers. In simple terms, we deploy a grid surrounding the single-line, so that the vehicles move within the grid depending on the specific users they are serving.

In our setting, we have another challenge that arises when analyzing scale for on-demand systems: which fleet to use. Most models that simulate ODRP assume a given fleet (as we describe further in Section 2.1). However, a proper scale analysis requires that the fleet is actually optimized, which is troublesome because the total fleet cannot be changed on-demand (only the operative fleet can be optimized). Here we propose a method to compute the fleet together with the assignment decisions, which is interpreted as *a posteriori*, i.e., it determines the fleet size that should have been used to serve the demand.

As we aim to lower congestion, the ODRP system we study follows rules that resemble public transport operations. It is non-profit, and the costs of all agents (users and operators) are considered when deciding how to assign vehicles to users. The routes of the vehicles are instructed by a central operator aiming to minimize a function that represents total social costs, where we impose that all users must be served. The use of the single-line model entails that the zone covered by the ODRP system represents what could be served by a single transit line. Moreover, we do not impose a door-to-door service, i.e., the system might decide (on-demand) pick-up and drop-off points that require some short walks if doing so improves the system's overall performance.

Our main contribution is to identify new sources of scale economies and diseconomies in ODRP systems that enlighten the potential and obstacles that need to be overcome for ODRP to succeed. Some of these sources are specific to ODRP systems, as they depend on how the flexible routes followed by the vehicles evolve when the number of passengers grows. Furthermore, we propose a way to compute the fleet size in ODRP together with the assignment decisions, which can be utilized for other types of analysis beyond the objectives of this paper. We also show the potential of relaxing the door-to-door scheme when all requests must be served, and compare our results with an idealized public transport system.

The paper is organized as follows. Section 2 revises relevant previous studies. Section 3 explains the methodology that we use to compute the ODRP system's fleet and to expand the single-line model. Section 4 shows the results of the numerical simulations. The most relevant qualitative conclusions from this paper, regarding scale effects and the circumstances that favor the use of ODRP, are described and mathematized in Section 5. Finally, Section 6 concludes and proposes some directions for further research.

## 2. Related works

### 2.1 Fleet sizing in on-demand ridepooling systems

Deciding which fleet to use in an ODRP system is not an easy task. Contrary to public transport, the routes cannot be known in advance, so the usual techniques dealing with cycle times and desired frequencies cannot be applied here. Such difficulties have been faced with different approaches that we now describe.

The most usual approach is to work with fleets of fixed size. In order to determine which one is optimal, or at least gain some intuition about that, it is habitual to repeat the same numerical experiments with different fleet sizes to analyze which size responds better to a given demand (Alonso-Mora et al., 2017, Levin et al., 2017, Lokhandwala & Cai, 2018, Wang et al., 2018). Other studies seek the minimal fleet able to meet some exogenous conditions on the quality of service. Daganzo & Ouyang (2019) and Martinez & Viegas (2017) require to serve all the demand, although the latter also compare the results obtained with larger fleets. Spieser et al. (2014) consider bounds on the number of passengers waiting to be served, and Fagnant & Kockelman (2018) aim at fulfilling some predefined waiting times.

Alternative rules to analyze fleet size in ONRP include the proposals of Santos & Xavier (2015), who assume that the number of vehicles has to be proportional to the number of requests, a rule that is obtained as a result by Kang & Levin (2021) when following an assignment policy that aims at maximizing the number of users per vehicle; Pinto et al. (2020), who assume the availability of a budget, shared with public transport, that has to be respected; and Fielbaum (2020), who makes a weighted optimization between users' and operators' costs under simplifying assumptions that lead to the prediction of exact fleet sizes. Cap & Alonso-Mora (2018) explain that the optimal fleet size also consider both types of costs and study the corresponding multi-objective problem, proposing a method to compute the Pareto front.

It is worth mentioning that our techniques for fleet sizing are inspired by Cap et al. (2021), who optimize the fleet together with the assignment decisions between vehicles and batches of users. The main difference is that once a group is created, they assume that it has to be wholly served before updating the vehicle's route, so that the main question is how to chain different groups to reduce the number of required vehicles. Such an approach extends the study by Vazifeh et al. (2018), who optimizes the fleet for a non-shared on-demand system. In both studies, the fleet is computed as if the demand was known, an approach that we also follow here.

### 2.2 The single-line model for public transport analysis

The single-line model refers to analyzing public transport systems by considering a line as isolated from the rest of the system. We now describe the scale effects that have been identified using this model. Most of such effects have been shown to remain valid for each line when a network is considered (Fielbaum et al., 2020a).

The stream of studies based on the single-line model was pioneered by Mohring (1972), who identified one of the main sources of scale economies in public transport (now known as the "Mohring Effect"): more passengers require more buses, which increases the service frequency and diminishes waiting times for everybody. His model was later extended by Jansson (1980) to consider optimal bus capacities and time at stops, where a source of diseconomies of scale emerges, namely that an increase in the number of users yields the utilization of larger buses, making users to spend more time waiting for other passengers to board and alight (an effect that can be compensated by changing the number of doors per vehicle as explained in the next paragraph). Evans & Morrison (1997) discovered yet another source of scale economies with an extension of this model: an increase in the number of users enables spending more resources in preventing accidents and disruptions in the service. These consecutive advances have been surveyed and expanded by Jara-Díaz & Gschwender (2003).

The single-line model has been used for other purposes (different than scale analysis) as well. Jara-Díaz et al. (2017, 2020) have studied the impact of accounting for two different periods in the optimal design of the frequencies and fleets for a single line; Hörcher & Graham (2018) have focused on a spatially unbalanced line; Basso et al. (2020) have studied the evolution of the urban structure surrounding a single line; Oldfield & Bly (1988) analyzed the optimal bus size; Jara-Díaz & Tirachini (2013) studied the optimal payment technology and number of doors per vehicle; and Tirachini & Antoniou (2020) analyzed the impact of utilizing automated vehicles in a public transport line. Finally, on a similar note, a sort-of single-model has also been used to study optimal spacing between parallel lines, by replicating the single line several times in space (Fielbaum et al., 2020b, Kocur & Hendrickson, 1982, Chang & Schonfeld, 1991). These topics are discussed at length by Hörcher & Tirachini (2021). As such, variations of the single-line model have been used for decades and are still used nowadays to improve our understanding of the structural aspects of public transport design.

## 2.3 Integrating ridepooling and public transport

The potential of a system that integrates ODRP services with traditional public transport lines has been acknowledged by several researchers in the past few years, although there is still no systematic way to model such an integrated system. Several authors assume an integration that is achieved by a feeder-trunk structure, in which the ODRP systems serves as the feeder (Banerjee et al., 2021, Chen & Nie, 2017, Chen et al., 2020, Fielbaum, 2020; Ma et al., 2019, Wen et al., 2018) or replacing traditional public transport lines in low-demand areas (Basciftci & Van Hentenryck, 2021, Kim & Schonfeld, 2014, Mahéo et al., 2017, Pinto et al., 2020, Shen et al., 2018). Also studying feeder systems, Mo et al. (2021) study what happens when ODRP and public transport compete rather than collaborate, while Fielbaum (2020) argues that if users do not share a common destination, the system design becomes inefficient due to the difficulty of finding passengers whose routes are compatible.

The intuition of utilizing flexible services when demand is low seems correct, as studied by some authors that compare fixed and flexible lines. Badia & Jenelius (2020), as well as Papanikolaou & Basbas (2020), have rested on specific functional forms that approximate the ODRP systems, finding that they should be preferred not only when the demand is low but also when the areas to be served are small, and trips are short. Li & Quadrifoglio (2010) and Quadrifoglio & Li (2009) use continuous approximation models and identify the discomfort of walking as another relevant parameter that determines which type of system should be preferred. Similarly, Calabrò et al. (2021) use microsimulation to find that flexible services are better in rural areas. On the contrary, Bischoff et al. (2019) suggest that public transport could be fully replaced by ODRP in small or medium cities, while Viergutz & Schmidt (2019) conclude that rural areas should use line-based on-demand services rather than completely flexible routes.

It should be noted that all these models assume that the flexible systems provide door-to-door service (or station-to-door, when it is solving the last-mile problem), which is a common assumption as most real-life on-demand systems operate in that way. However, operating door-to-door is not mandatory for this type of system. Actually, previous research has consistently shown that requesting some users to walk either to personalized pick-up and drop-off points (Fielbaum, 2021, Fielbaum et al., 2021) or to group meeting points ((Bischoff et al., 2019, Li et al., 2016, Li et al., 2018, Stiglic et al., 2015)) can enhance ODRP services significantly. Such ideas are already applied in real life: the shared-mobility platform Jetty in Mexico City asks passengers to be at specific pick-up points to be able to board a shared car or van; and users can monitor the location of the vehicle in real-time before boarding (Tirachini et al., 2020).

## 3. Methodology

This Section explains the methodology we develop to face the two main issues hindering scale analysis for ODRP. First (Section 3.1), we extend the method from Fielbaum et al. (2021) (which assigns users to vehicles allowing for some walks) in order to compute which fleet to use endogenously. In general terms, this is done by having a potential vehicle for each upcoming request, so that the vehicle will actually be utilized only when paying its capital cost is more efficient than using the previously available fleet. Second (Section 3.2), we extend the single-line model by deploying a grid surrounding it. This grid might be seen as implicit when such a model is used to study public transport systems, but requires to be explicit for ODRP as it enables flexible routes, so that we can study the evolution of such routes when the demand grows. Finally (Section 3.3), we explain how we determine some bounds on the level of service for ODRP, inspired by what users actually experience in public transport.

### 3.1 Computation of the number of vehicles in the ODRP system

In order to compute the fleet size together with the assignments between vehicles and users, we build upon the ODRP model proposed by Fielbaum et al. (2021). Such a model extends the one by Alonso-Mora et al. (2017) by optimizing the pick-up and drop-off points, which might differ from the actual origins and destinations of the users when asking them to walk increases overall efficiency. Both models determine how to operate a fixed fleet of vehicles to serve the emerging requests. We extend these works by computing the fleet endogenously. We first explain briefly how the original methods work, and then describe this extension.

The ODRP system operates over a directed graph $G = (N, A)$. Each request $r = (o_r, d_r, t_r)$ is a triplet, representing the origin, the destination, and the time in which the trip is requested. Both the origins and the destinations are assumed to be placed over the nodes of the graph. The assignment model works using a *receding horizon* approach, meaning that it accumulates the requests that emerge during a fixed amount of time $\delta$ and assigns them all at once, which updates each vehicle's route. When such an assignment is decided, the vehicles follow their updated routes, and the system begins to accumulate requests for a time $\delta$ again, starting a new iteration.

Let us focus now on a single iteration, denoting by $R$ the set of requests to be assigned, and by $V$ the current state of the fleet of vehicles. Each vehicle is characterized by its position $P_v$ and the set of requests assigned to it $S_v$ (either in the vehicle or waiting for it). The assignment between $R$ and $V$ takes place following these three steps:

- Determine which are the feasible *trips*. A trip $T$ is defined by a group of requests $req(T) \subseteq R$ and a vehicle $veh(T)$, so that $T$ is feasible if the requests in $req(T)$ can be transported together by $veh(T)$, respecting some bounds on waiting and walking times, and on total *delay* (denoted, respectively, $\Omega_w, \Omega_a$ and $\Omega_d$). Such bounds affect users in $req(T)$ and in $S_{veh(T)}$. The delay is defined as the extra time faced by a user compared to beginning her trip immediately, with no walking and following the shortest path between her origin and destination. Each trip $T$ might be served by more than one route so that taking the route $\pi$ imposes a cost to the system given by Eq. (1):

$$cost(T, \pi) = \sum_{r \in req(T)} c_U(r, T, \pi) + \sum_{r \in S_{veh(T)}} \Delta c_U(r, T, \pi) + \Delta c_O(\pi) \qquad (1)$$

    Where the first term represents the users' costs for passengers in trip $T$, defined as a weighted sum between waiting, walking, and in-vehicle times; the second term represents the extra costs induced to the users that were being served by the vehicle prior to this assignment (because their waiting and in-vehicle times can increase); and the third term expresses the increase in operational costs, that are assumed to be proportional to the route length. The route that offers the minimum cost is selected, so that the trip $T$ is characterized by a single figure $cost(T)$.

    It is worth commenting that computing all the feasible trips can be computationally expensive, as their amount can increase exponentially with the number of requests. Such an issue is faced first by making a smart search of the feasible trips (using that if vehicle $w$ is able to serve group $G$, then it must be true that $w$ can serve every subset of $G$ as well), and also by using a number of heuristics, explained in detail by Fielbaum et al. (2021), to compute the sequence in which the users are served and the pick-up and drop-off points.

- Once the set $\Gamma$ of potential trips is known with their respective costs, some of them are selected and constitute the actual assignment. To do this, an Integer Linear Programing (ILP) problem defined by Eqs. (2)-(4) is solved:

$$\min_{x,z \in \{0,1\}} \sum_{T \in \Gamma} x_T cost(T) + \sum_{r \in R} p_{KO} z_r \tag{2}$$

$$\text{s.t.} \quad z_r + \sum_{T: r \in req(T)} x_T = 1 \quad \forall r \in R \tag{3}$$

$$\sum_{T: veh(T)=v} x_T \leq 1 \quad \forall v \in V \tag{4}$$

Binary variables $x_T$ represent the trips that are going to be executed (marked by $x_T = 1$). It is not always possible to serve all the trips (the number of vehicles might not be enough), so rejected requests are marked by $z_r = 1$. Each rejected request imposes a penalty $p_{KO}$ to the system, so Eq. (2) is the objective function to be minimized when deciding the assignment. Eq. (3) ensures that each request is either rejected or belongs to a trip that is going to be executed, while Eq. (4) ensures that each vehicle is assigned to no more than one trip.

- Finally, a rebalancing step instructs idle vehicles (i.e., those with no requests before the assignment and did not receive anyone here) to move to certain areas where more vehicles are needed. We do not explain the details here because we do not use such a procedure in this paper. We do execute a simple rebalancing step when modeling a feeder model, which we explain in Section 3.2

In this paper, we extend this model to decide how many vehicles to use at the same time we decide the vehicle and user assignments. To do so, we assume that the system begins with no vehicles, and that there are some spots in the city (which is a set of nodes $M \subset N$) where potential vehicles are placed. At each iteration (i.e., each time a batch of requests is assigned), the fleet of vehicles is composed of two sets: the one inherited from the previous iteration, plus a set containing one *non-activated* vehicle per request $r \in R$, that is located in the node in $M$ that is closest to its origin $o_r$. If a non-activated vehicle is assigned to a group of requests, an activation cost $c_A$ has to be paid, and the vehicle becomes available for the rest of the period of operation without paying $c_A$ again. This is formalized by altering the cost of the trips. Denoting by $A(v) = 1$ if vehicle $v$ is activated (i.e., inherited from a past iteration) and $A(v) = 0$ if not, Eq. (1) is modified to build the new cost function $cost_A(T)$, given by

$$cost_A(T) = cost(T) + c_A \cdot [1 - A(veh(T))] \tag{5}$$

These potential vehicles are interpreted as optimizing *a posteriori*, and $c_A$ should include all the costs that do not depend on the distance driven by the vehicle, such as capital costs. The fleet optimization cannot be done online, so optimizing a posteriori (i.e., as we knew all the requests) is appropriate. In this case, the receding horizon approach is not meant to represent the online optimization, but a way to manage the impossibility (due to its enormous complexity) of assigning the whole set of requests optimally at once. An alternative interpretation is that the fleet of vehicles is actually available, and the activation costs refer to hiring a driver for that day, which would require that $c_A$ captures only her daily wage.

Following Tirachini & Hensher (2011) and Jara-Díaz et al. (2017), we assume that both components of operators' costs grow linearly with the capacity of the vehicle. That is, if we denote by $K$ the number of users that can use a vehicle simultaneously, and by $c_O$ the proportionality constant that defines the costs depending on the routes' lengths, then:

$$c_O = c_{O1} + c_{O2}K, \quad c_A = c_{A1} + c_{A2}K \tag{6}$$

As we aim at designing a system that could be integrated into public transport, we must have zero rejections. As we now have one non-activated vehicle per request, it is always feasible to serve everybody. Therefore, we do not longer include variables $z_r$ in the ILP to be solved, and we modify Eq. (3) accordingly to ensure that each request belongs to exactly one assigned trip, i.e.

$$\sum_{T:\, r \in req(T)} x_T = 1 \quad \forall r \in R \tag{7}$$

Finally, we include yet another extension to the base model: we assume that a fixed time $\tau$ is spent each time the vehicle stops to pick up or drop off one or more passengers. We include this fact because it is relevant when analyzing scale economies, as sometimes the vehicle might use a single stop for more than one pick-up/drop-off, saving some time.

## 3.2 Extending the single-line model

The traditional single-line model studies the operational characteristics of a public transport system in which the vehicles follow a predefined path, so everything is one-dimensional. Specific versions are:
- The circular model, in which the line tours a circuit that presents the same average number of users at every point. This model represents a line that carries a similar load all along its way.
- The linear model, in which vehicles travel in both directions along a linear corridor between two terminals. A particular case of the linear model is the feeder model, in which users board the vehicle across its path, and they all alight at the end. This model represents a line that goes to some relevant final destination, typically a transit station, to board a high-capacity public transport mode (e.g., rail, Bus Rapid Transit).

In any of these alternatives, the vehicle route is fixed beforehand and always the same. We aim to extend this model, keeping most of its simplifying assumptions that make it a powerful tool, but allowing for online decisions regarding the routes. To do that, we deploy a grid surrounding each bus stop, where exact origins and destinations are situated. In the traditional model, such a grid can be seen as an underlying street pattern, that does not need to be explicit because users need to walk towards the (fixed) bus stops anyhow. In such a case, walking times and distances are assumed exogenous, meaning that the operation and optimization of the public transport line are not affected.

To be precise, we assume that each bus stop belongs to a *zone*, which is an $a \times b$ grid, with $a, b$ odd numbers, so that the bus stop is located at the center of the grid. That set of stops represent where the potential vehicles for the ODRP system are located (the set $M$ defined above). The central streets of the grid are bidirectional, and vehicles tour them at velocity $v_1$, whereas the rest of the streets are unidirectional[1], with alternate directions and velocity $v_2$, whith $v_2 < v_1$. Having streets of different velocities and directions help to capture that not all routes are equally good for the vehicle to follow. The whole network is formed by chaining consecutive zones. If there are $Z$ zones, this makes a $Z \cdot a \times b$ grid in the feeder model; in the circular model, the same happens, but the last zone is chained with the first one, forming a circular grid. Both networks are depicted in Figure 3.1.

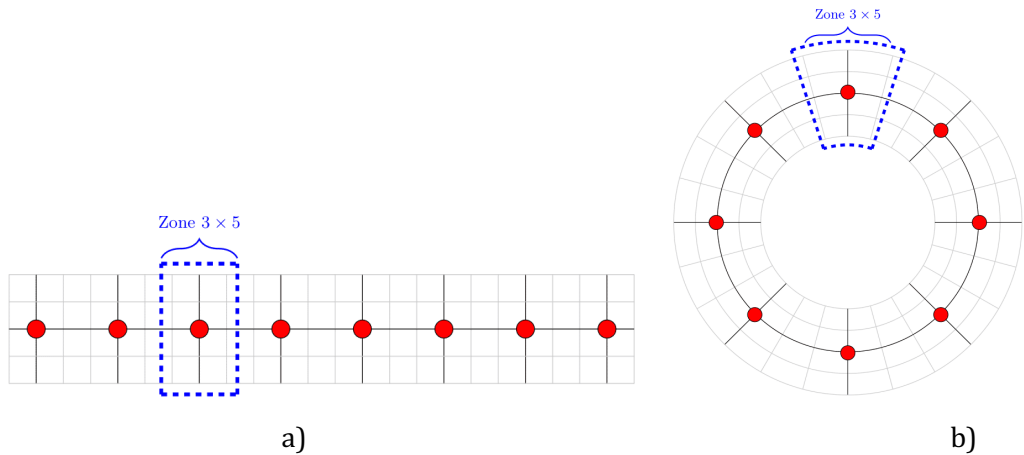

**Figure 3.1** Extensions of the single-line model to recreate the network in which the ODRP system operates, replacing either a feeder line (a) or a circular line (b). Origins can be placed in any intersection, and the same happens with destinations in the circular model. In both cases, there are 8 zones, each formed by a 3x5 grid. Red dots represent the stations in which the ODRP vehicles begin their journeys. Dark black streets are bidirectional and can be toured with a higher speed. The traditional single-line model is recovered by considering only the long avenue that connects all the red dots.

Regarding the demand, we want to keep the homogeneity assumptions from the single-line model but enabling for more complex routes. A constant number of users $Y$ emerge per time unit, and the exact origin is random: we first choose the zone with uniform probability; within that zone, the central node is chosen with probability $p$, the rest of the nodes located in the central streets with probability $p\gamma$, and the nodes out of the central streets with probability $p\gamma^2$. The parameter $p$ is adjusted to make the sum of the probabilities within every zone equal to 1, and the parameter $\gamma \in (0, 1)$ controls how dispersed the demand is within a zone (the lower the $\gamma$, the more concentrated the demand in the vicinity of the bus stop). The destination is computed differently depending on the model: in the feeder one, everybody goes to the center of the final zone, whereas in the circular model, the destination zone is located $l$ zones ahead, plus a random variable that is obtained rounding a normal distribution with mean zero and variance $\sigma^2$; the exact destination is found within that zone using the same rules involving $p$ and $\gamma$ as for the origin.

---
[1] In the feeder model, the first and last transversal streets are also bidirectional so that there are no isolated nodes.

As mentioned above, in the feeder model we need to rebalance idle vehicles to prevent them from accumulating in the common destination of all users: after reaching that node, they are sent towards the central node of the first zone (i.e., the one located at the largest distance from the shared destination). Such vehicles will not necessarily arrive there because they will be considered available in the following iterations, meaning that they might receive new passengers before reaching the first zone.

## 3.3 Definition of the bounds in the quality of service

As explained above, the assignment procedure in ODRP imposes predefined bounds on the quality of service, namely maximum waiting ($\Omega_w$) and walking ($\Omega_a$) times, as well as a maximum total delay[2] ($\Omega_d$). Defining such bounds is a relevant issue, as it has relevant impacts on the performance of the ODRP system. For instance, if the bounds are too tight and users are too spread, then the system might require to allocate almost one different vehicle per request, leading to a huge fleet; on the other hand, if the bounds are too large (or inexistent), one single vehicle might be able to serve all the requests, but offering an awful (and unrealistic) quality of service.

We adapt the bounds depending on the total number of users. That is, for each demand level, we compute the corresponding bound, which does not vary in time (i.e., it is calculated offline). Note that this mimics what passengers usually face when using public transport: when they want to make a trip on a high-demand corridor, they can rapidly find a bus (or any alternative mode they are using), and the contrary happens in low-demand areas. Thus, we define the bounds to replicate this behavior, by means of the classical single-line model by Jansson (1980) and the posterior adaptations by Jara-Díaz & Gschwender (2009), described in Appendix A.1, where the key variable is the optimal frequency $f$. The bounds are defined as follows:

- **Waiting:** The maximum waiting time that can be faced in the public transport system occurs when a passenger arrives at the station just after a bus leaves, waiting for $1/f$ (a quantity that decreases with the number of passengers). Recalling that when a vehicle is activated, it goes from the station to the pick-up point, we need to ensure that there is always enough time to wait for such a movement. Denoting by $t_1$ the vehicle-time from the station to the corner of the zone's grid, we use $\Omega_w = \max\{\frac{1}{f}, t_1\}$.
- **Walking:** The maximum amount of walking in the public transport systems is $t_2$, defined as the walking time between the station and a corner of the zone's grid, so we use $\Omega_w = t_2$.

---

[2] Such bounds ensure that users will indeed accept the assignment proposed by the system rather than searching for an alternative mode. Moreover, without them the algorithmic burden of the problem would be unmanageable, as every possible group of users could be feasibly served by any vehicle.

When we simulate the case in which ODRP offers a door-to-door service, this bound is reduced to zero.
- **Delay:** There are two sources of delay in public transport with respect to the time in the vehicle: walking and waiting. The first one should be accounted for twice, at the origin and destination. Therefore, we use $\Omega_d = \max\{\frac{1}{f}, t_1\} + 2t_2$.

## 4. Results

We simulate one hour of operation of the ODRP system, for increasing demand levels, in order to identify scale effects. The numeric values of the parameters are shown in Table A.1 in the Appendix. All figures in this section use a logarithmic scale in the x-axis, because the phenomena that we study tend to stabilize when the number of passengers is high, so zooming in the lower values helps the analysis. The simulations are run for different sizes of the ODRP's vehicles, including vehicles with capacity for 2, 3, 4, and 5 passengers.

Most results consider the base case in which we assume the availability of automated vehicles (AV) and walks are allowed. We assume that AV differ from human-driven vehicles in the parameters that represent operators' costs (because there are no wages, but such vehicles might be more expensive). In particular, in our analyses, the velocities at which vehicles run do not depend on such a technology: differences in velocity are still uncertain, as AV might run faster (thanks to better coordination among vehicles) or slower (due to safety reasons), and a specific assumption on this may have a large impact on the results (Tirachini & Antoniou, 2020).

### 4.1 Circular model

Results of the circular model are exhibited in Figures 4.1 to 4.5. Figure 4.1 shows a condensed way to describe the quality of service of the ODRP system from the users' point of view: total delay, i.e., the extra time faced by them when they use this system instead of traveling in a private vehicle. This total delay includes walking time, waiting time, and detour once on the vehicle. The relationship between total delay per passenger and demand, which influences the existence of scale economies, is clear. Scale effects are remarkable: At the very beginning of the curve, up to around 250 passengers/h, there is a reduction of total delay. However, when the number of passengers continues to grow, diseconomies of scale appear as average delay increases to 5 min/passenger for demands up to almost 1000 passengers/h. Then, the average delay is once again reduced, to reach around 2 min/passenger for 3000 passengers/h.

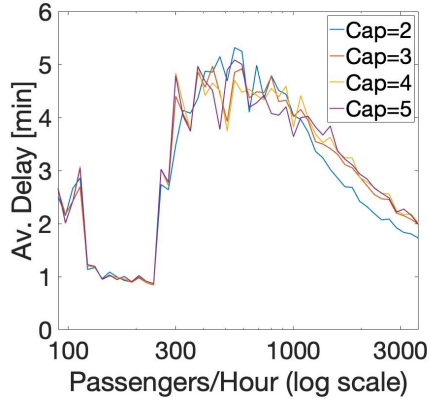

**Figure 4.1** Average total delay faced by the users of the ODRP system in the circular model, as the number of hourly passengers grows. Different curves represent different vehicles' sizes.

To understand the curves from Figure 4.1, we disentangle the total delay per passenger in its three components in Figure 4.2: Waiting (a), walking (b), and detour (c). Waiting times evolve similar to total delay. Let us begin our analysis after the strong drop at the beginning of the graph. The remaining of the curves reflect that average delay first increases and then slowly decreases.

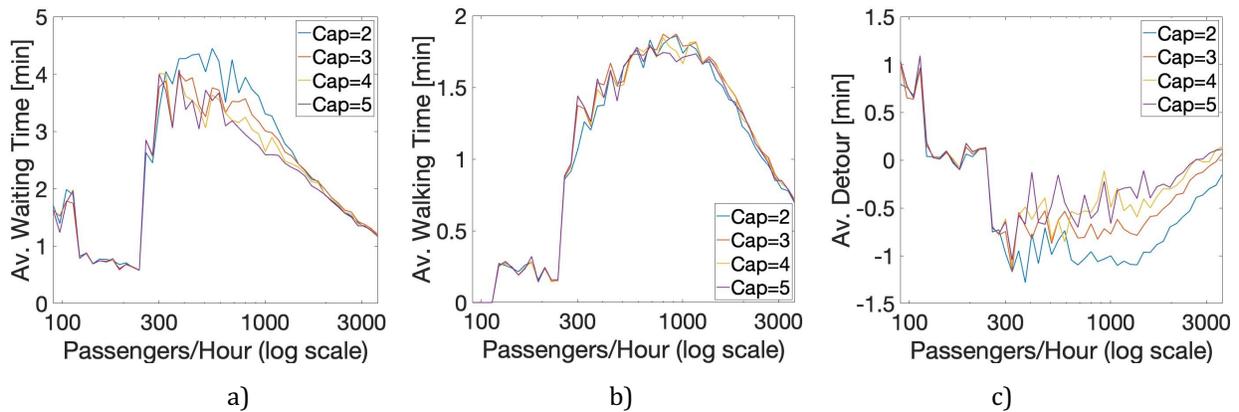

**Figure 4.2** Average waiting time (a), walking time (b) and detour (c), faced by the users of the ODRP system in the circular model, as the number of hourly passengers grows. Different curves represent different vehicles' sizes.

Diseconomies of scale emerge when $Y$ reaches about 250 passengers/h. Until that point, the system operates almost as a private service, i.e., there is little sharing because it is difficult to find compatible users, implying that most users travel alone[3]. When vehicles begin to be shared with more people, one of its consequences is that vehicles do not go directly to pick up the users but deviate to serve some co-travelers, hence increasing waiting times. This effect dominates for demands greater than 250 passengers/h. The same phenomenon can be seen related to walking and the detour, which also start to increase when crossing the same threshold, and is verified in Figure 4.3, where we show the average load of the vehicles at the end of the simulation, revealing that the load begins to grow at the exact same threshold. Actually, there is almost no walking at all when the demand is very little. Noteworthy is that the smaller the vehicle, the lower the detour, and that detours can be negative when there is some walking because the distance between the pick-up and drop-off points might be lower than between the corresponding origins and destinations. Therefore, we have identified **a relevant source of diseconomies of scale in ODRP systems: an increase in the number of users implies that the vehicles will be shared by more passengers, which increases average traveling times.**

---

[3] This occurs in some real-life scenarios. In Fürstenfeldbruck, Germany, during some specific time windows in which the demand is very low, the public transport agency sends private taxis to fulfill it.
See https://www.mvv-muenchen.de/mobilitaetsangebote/mvv-ruftaxi/index.html (Accessed: 19/05/2021).

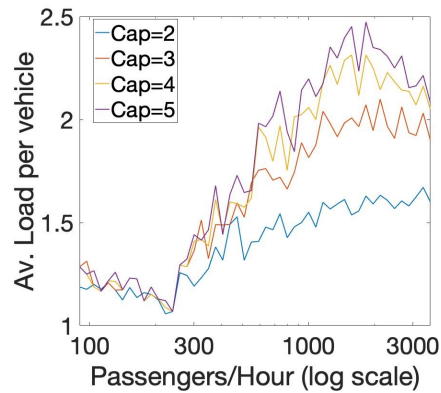

**Figure 4.3** Average number of users in the vehicles at the end of the simulation of the ODRP system in the circular model, as the number of hourly passengers grows. Different curves represent different vehicles' sizes.

As discussed by Fielbaum & Alonso-Mora (2020), the fact that routes are not known beforehand, but depend on the travelers, is specific to models that are both shared (otherwise vehicles follow shortest paths) and on-demand (otherwise vehicles follow fixed routes). Therefore, this source of scale diseconomies is specific to ODRP mobility systems. We denote this source as the "Flex-route Effect".

In this model, the Flex-route Effect increases waiting, walking, and in-vehicle times as vehicles increase their load. It is noteworthy that there are other negative externalities that the model does not directly capture:
- Fielbaum & Alonso-Mora (2020) identify two types of **unreliability:** The "one-time unreliability", defined as changes that take place while a trip is executed due to emerging requests, and the "daily unreliability", that refers to facing different conditions each time a trip is repeated. Both types of unreliability worsen when vehicles are more loaded, i.e., the Flex-route Effect increases unreliability. It is worth mentioning that this is not a minor issue: for instance, Alonso-González et al. (2020) have estimated the value of reliability (that deals with the daily unreliability discussed above) to be approximately a half of the value of time.
- Sharing the vehicle with more users can be uncomfortable by itself, as studied by Ho et al. (2018), König & Grippenkoven (2020), and Lavieri & Bhat (2019), who propose the so-called "**willingness to share**" to study the difference in comfort between traveling alone or with other users. Note that this effect only occurs when vehicles start to increase their load (namely, when they pass from one to two passengers). However, it is complemented afterward with the increase in **crowding**, i.e., the discomfort because there is less space in the vehicle, which has been thoroughly studied in traditional public transport systems and surveyed by Tirachini et al. (2013).

The Flex-route Effect eventually gets exhausted. At some point, the vehicles no longer increase their load (when they are running at capacity, considering their current passengers and the ones that are waiting to be picked up). When this happens, Figures 4.2a and 4.2b reveal that waiting and walking times begin to diminish. Regarding waiting times, this is the analogous version of the Mohring Effect for this type of mobility system: an increase in the number of requests is satisfied with an increase in the fleet size; therefore, it is more likely that an available vehicle is nearby. The explanation for the reduction in walking times is similar, and it has also been recognized as a "spatial counterpart of the Mohring Effect" by Fielbaum et al. (2020b): more vehicles also means that they are denser in space.

Additionally, more users imply that it is possible to find better matching among them, which also reduces waiting and walking times. When the demand is greater, origins and destinations are also denser in space, meaning that vehicles require shorter detours to combine compatible passengers[4]. We can synthesize these scale effects by stating that **two relevant sources of scale economies in ODRP are that the increase in the number of users leads to 1) a larger fleet, which reduces waiting and walking times, similar to the Mohring Effect in fixed-route public transport, and 2) match users whose routes are more compatible.** For the sake of simplicity, let us denote the first phenomenon again as "Mohring effect". The second source will be referred to as "Better-matching Effect", and is the ODRP analogous to the fact that more direct lines can be offered in public transport as the demand grows (a source of scale economies identified by Fielbaum et al., 2020a).

The quick drop at the beginning of the curve is explained by the Mohring Effect, but only regarding waiting times. As vehicles' load does not increase yet (and actually might decrease slightly, due to the decrease in $\Omega_w$ and $\Omega_d$), neither the Flex-route nor the Better-matching Effects operate significantly. Moreover, the Mohring Effect is usually more important at low demands because when the number of vehicles is already large, the marginal impact of an additional vehicle is low in reducing waiting times.

It is worth noting that the three phenomena identified so far act on top of each other, counterbalancing their impacts. That is to say, the positive scale effects (Mohring and Better-matching) operate even when the total delay presents global diseconomies of scale, but the other effect (more people sharing the vehicle) prevails, and vice-versa.

---

[4] Such an effect has also been observed in a real-life carpooling system (Scoop), where a private driver shares her trip with riders that emerge on-demand (Lehe et al., 2021).

The comparison among different vehicle sizes is also informative. The smaller the vehicle, the lower the load and thus the detour. Walking times are not affected significantly by the vehicle capacity adopted. On the other hand, waiting times are slightly larger for smaller vehicles when the demand lies in the range of 250-1000 passengers/h. As the fleet size is mostly unaffected by the vehicle capacity within that range (see Figure 4.4), it is more likely that the assigned vehicle is not immediately available when vehicles are small.

The evolution of operators' costs (depicted in Figure 4.4) is mostly characterized by scale economies when exceeding the threshold in which vehicles start to be shared more intensively (before the threshold, it exhibits an irregular pattern in which the randomness of the requests play the most relevant role). This is reflected in the fleet size (Figure 4.4 left), which also exhibits scale economies in public transport, but also in operating costs (vehicle hours traveled VHT, Figure 4.4 right). It is noteworthy that using smaller vehicles requires a larger fleet when used at capacity, which also increases VHT. Both curves eventually stabilize, meaning that this source of scale economies gets exhausted.

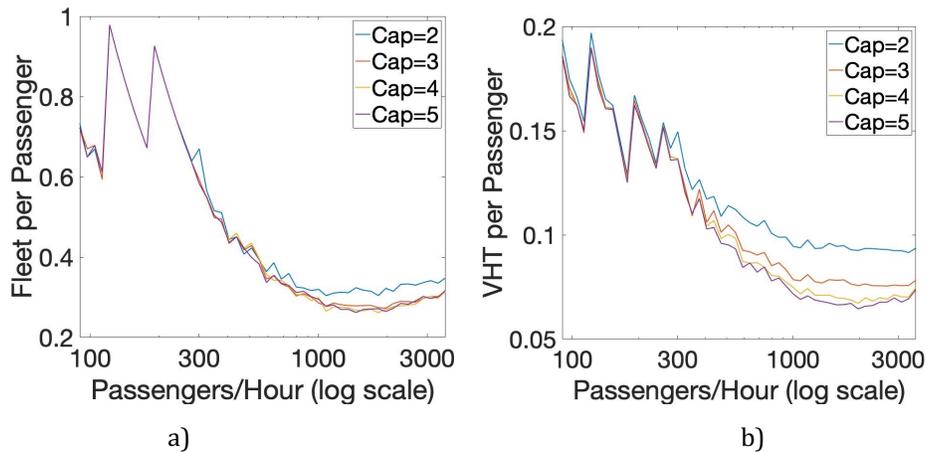

**Figure 4.4** Fleet size (a) and Vehicle-Hours-Traveled (b), normalized by the number of passengers, as this last quantity grows. Different curves represent different vehicles' sizes.

So far, we have exhibited results for a range of vehicle capacities, from 2 to 5 passengers/veh. However, the system should utilize vehicle sizes that minimize total costs. Our results indicate that the smallest vehicles (capacity 2) should be used if $Y \leq 550$, and capacity 3 thereafter. Figure 4.5 synthesizes scale effects for users and operators when the capacity is optimized. The delay curve (Figure 4.5a) looks almost exactly as Figure 4.1, meaning that all the scale phenomena discussed above remain valid. Figure 4.4 implies that both the number of vehicles and VHT still exhibit scale economies when the capacity is optimized, but there remains one aspect to be analyzed: the number of seats $S$, defined as the product of the number of vehicles and its capacities. Recall that, according to Eq. (6), operators' capital and operating costs depend both on the total number of vehicles and on $S$. The evolution of $S$ when the capacity is optimized is shown in Figure 4.5b: it is similar to what we observed regarding fleet size (first erratic and then scale economies), but with a small jump when the optimal capacity switches from 2 to 3 (around 600 passengers/h in Figure 4.5b).

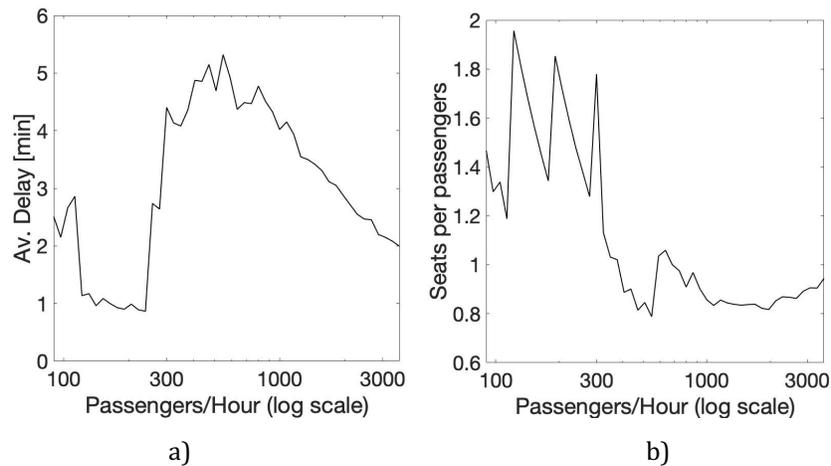

a)                    b)

**Figure 4.5** Average delay (a) and Seats per passenger (b), yielded by the ODRP system in the circular model, as the number of hourly passengers grows, when the optimal capacity is selected.

In Figure 4.6, we synthesize the results, including two alternative scenarios: forbidding walks (i.e., providing door-to-door service), and utilizing human-driven vehicles instead of AV, which diminishes capital costs but includes the drivers' wages. Figure 4.6 shows the average cost per user: in all three scenarios, we observe the same situation, namely, no clear trends for very low demands and economies of scale after a certain demand threshold is reached. This implies that the sources of scale diseconomies that we identified for the users get outweighed by the sources of scale economies for the operators, **leading to a global situation of scale economies that eventually get exhausted**. The comparison between the different scenarios and vehicle technologies also implies relevant conclusions:

- Using AV reduces the total cost to a considerable extent. This fits intuition, as having drivers for each small vehicle can increase global costs significantly (Bösch et al., 2018).
- However, when the number of users is large, enabling walks can be as important as changing the technology: both non-solid curves exhibit similar values of average total cost in Figure 4.6. **In fact, an ODRP system with human-driven vehicles that enables walking has a lower total cost than a system with AVs without walks, for some demand levels**. This is a remarkable finding regarding the value of designing an ODRP system with short walks.
- On the other hand, as there is little walking when the number of users is low (the system works similar to a private door-to-door service), for demands below 250 passengers/h the corresponding impact of enabling walks is negligible.

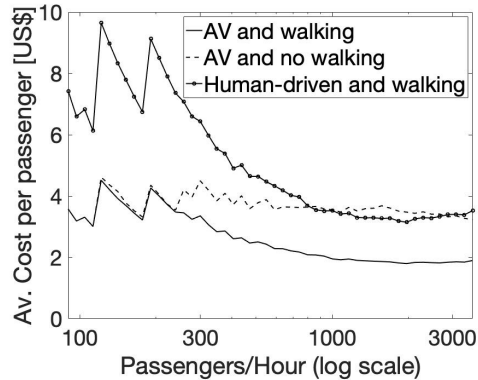

**Figure 4.6** ODRP's average costs in the circular model, as the number of hourly passengers grows, when the optimal capacity is selected. Different curves represent different types of vehicles and whether walks are enabled in ODRP.

## 4.2 Feeder model

As discussed in Section 2, much of the previous research has assumed that the ODRP services can help to solve the so-called "last-mile problem", i.e., as a feeder that connects the main transit stations with the specific origins (or destinations) of the users. For the ODRP system, the main difference is that everybody shares one extreme of the trip, which means that this model can also represent the case in which there is a very attractive destination, such as the city center. In our simulations, all users are traveling to the same destination (for instance, to take a second vehicle that does not affect the ODRP operation). Therefore, compatible routes are much easier to find. The only requirement is that when a vehicle is following a route, new passengers have to be located close to that route. This demand pattern has a significant effect in the simulations: for the same number of users, the number of feasible trips is multiplied by about twenty compared to the circular model. This increases the computational burden significantly, which is why here we simulate only up to capacities equal to four.

There is yet another relevant difference related to idle capacity. As users move all in the same direction, and the network is no longer circular, the vehicles must actively return in order to find some new passengers. Recall that this is executed through a rebalancing step: idle vehicles are sent towards the other extreme of the network, but they might not arrive there because they are still considered available for the emerging users.

The results of the simulation are depicted in Figures 4.7, considering the base model (AVs and enabling walks). Figure 4.7a condenses the information regarding users' costs by displaying the average delay, which shows the same trends as observed in the circular model, verifying the presence of the three sources of scale economies discussed above. Figure 4.7b shows the average load per vehicle (excluding vehicles being rebalanced), confirming that vehicles start to increase their load when some threshold in the number of passengers is exceeded. Moreover, the usage of the vehicles is much higher than in the circular model, and it is also higher than half the capacity, which is the expected load in public transport (because vehicles begin empty and get full along the way). When looking into total costs (Figure 4.7c), the same conclusions obtained for the circular model remain valid: average costs do not show a clear trend at the beginning, and scale economies prevail afterward until they eventually get exhausted.

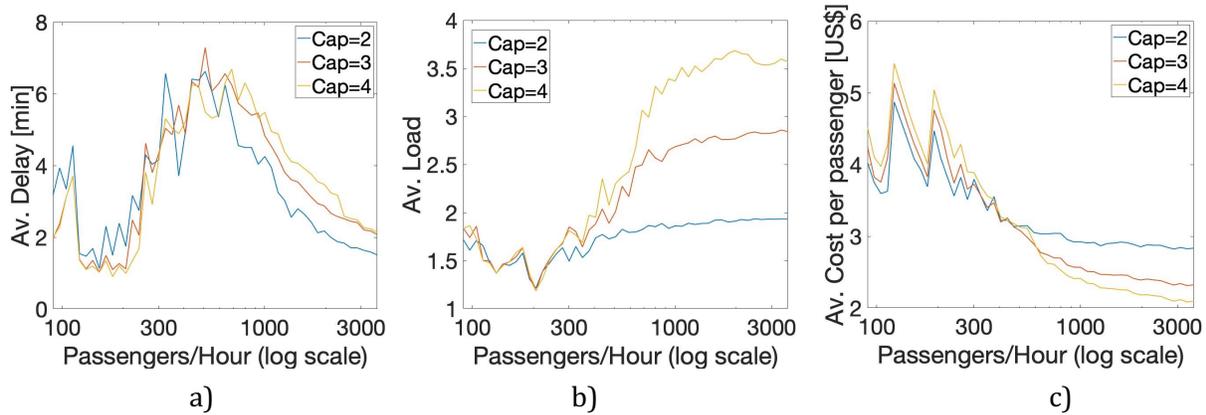

**Figure 4.7** Average delay (a), active vehicle's load at the end of the operation (b) and costs (c), faced by the users of the ODRP system in the feeder model, as the number of hourly passengers grows. Different curves represent different vehicles' sizes.

## 4.3 Comparison with an idealized public transport model

As the single-line model resembles the operation of a traditional public transport line, it is natural to analyze under which conditions ODRP could replace such a line. A precise model of the public transport is out of the scope of this paper; however, we do perform a comparison with an idealized public transport line, whose frequency and bus capacity are optimized following a procedure described in Appendix A.1. Such a comparison is depicted in Figure 4.8, and is informative regarding the trends in the respective curves. Figure 4.8 depicts the ratio between the total costs (including operators and users) of ODRP and public transport, considering both the circular and the feeder models. In ODRP, we select the capacity of the vehicles that minimizes total costs. ODRP is in the numerator, so that a value lower than 1 implies that ODPR provides the lowest total cost. The most relevant conclusions of this comparison are the following:

- ODRP should only be preferred if the demand is very low, in line with the findings of previous research efforts, as described in Section 2. This result is driven by the small size of the ODRP vehicles, and relates to the almost door-to-door scheme that results in such scenarios. This last characteristic also explains why ODRP is more competitive in the circular model for low levels of demand, as in the feeder model, public transport also has zero walking at the destination, softening the benefits of ODRP.
- For large demand levels, ODRP is more competitive in the feeder model. Note that in public transport, vehicles also need to "rebalance", i.e., to return empty to the other extreme of the network. In this case, all vehicles have to arrive there, as their route is fixed[5]. In ODRP, they do not need to arrive at that extreme, so that flexibility plays a role in diminishing the idle capacity of the system.
- For large demand levels, curves tend to stabilize, which is a natural result of the constant returns to scale that characterizes all these systems in such scenarios.

In all, if one has to choose between using only ODRP (with small vehicles) or only traditional public transport, the former should be chosen only for low-demand zones. However, our results regarding the presence of scale economies when the demand is large, suggest that other types of integration could yield even better results, utilizing both systems in some complementary way to take advantage of the good quality of service that can be offered to the users. How to design such an integrated system is a broad question that goes beyond the scope of this paper, but recognizing that there might be room for improving public transport provision in high-demand zones by means of smart utilization of ODRP systems is a promising venue for further inquiry.

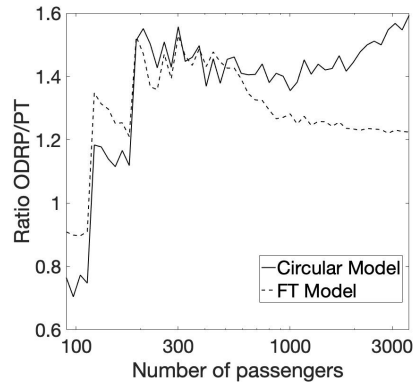

**Figure 4.8** Comparison between ODRP and public transport average costs as the number of hourly passengers grows, when the optimal capacity is selected, using AVs and enabling walks. Different curves represent the circular and the feeder model.

## 5. Scale analysis in ridepooling: A synthesis

ODRP services present sources of economies and diseconomies of scale. The former ones prevail when all costs are accounted for. However, when looking into users' costs only, there might be situations in which the negative externalities govern the system. To be more precise, we have identified three sources for scale analysis (equivalently, three types of externalities caused by new users). They all emerge from the analysis of users' delay (Figures 4.1 and 4.7a), how this is composed by waiting, walking, and detour (Figure 4.2), and the increase in vehicles' load when the demand grows (Figures 4.3 and 4.7b).

- **The Mohring Effect,** operating exactly as in public transport: new users induce the utilization of a larger fleet, which reduces average waiting and walking times.

---

[5] Both problems might be faced with ad-hoc techniques like having some vehicles serving only the last portion of the line, i.e., a "short-turning" strategy, potentially combined with deadheading, as studied by Cortés et al. (2011).

- **The Flex-route Effect:** This phenomenon is defined as the degradation in the perceived quality of service due to the increased vehicle load as a response to a greater demand. Intuitively, as the vehicle routes are not defined a priori but adapted to the specific users being served, the quality of service perceived by the users is sensitive to the route choice. When the demand grows, it is more likely to find users who can share the vehicles, eventually increasing their load. This increases the detours required by the system, which in turn increments waiting times. Moreover, the chance of walking instead of having a door-to-door service increases as well, because the time savings from walking are larger when more other passengers are affected. This effect is illustrated in Figure 5.1, where we show how the blue passenger increases all the components of her travel time when the vehicle serves a new user. The Flex-route Effect gets exhausted when vehicles run at capacity (or almost). It can be interpreted as similar to a well-known fact in public transport, namely that new users increase vehicle load, which in turn increases the time spent at stops waiting for boarding and alighting passengers. However, in ODRP, the route itself gets affected, so that the effect can be much more significant.

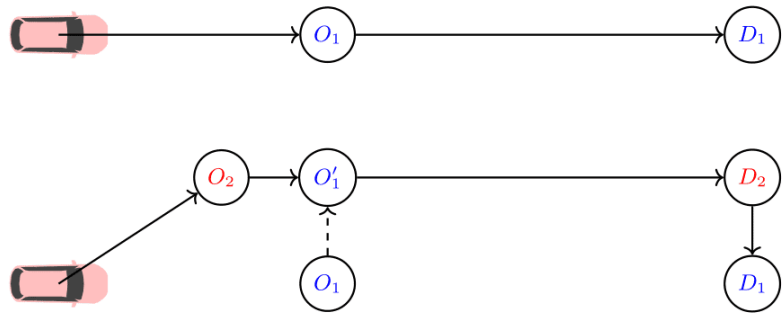

**Figure 5.1** Example of the Flex-route Effect. The number of passengers is low in the top row, so users do not share the vehicle, and the blue passenger faces little waiting time and no detour. When the demand grows (bottom row), a new red co-traveler appears close to her, which increases her waiting time, requires her to walk (marked with a dotted arrow), and implies a detour, degrading her perceived quality of service.

- **The Better-matching Effect:** It is defined as the ability to create groups whose routes are more compatible with each other when the demand grows, thanks to a larger pool of requests to choose from. Intuitively, as new users enter the system, it is more likely to find users whose origins and destinations can be matched without long detours. This can be seen mathematically by noting that the optimization problem described by Eqs. (2)-(4) (or the equivalent step in the corresponding assignment algorithm) exhibits a larger feasible set with greater demands, which always leads to better results. This effect manifests in a clearer way when vehicles load does not increase, as then the systems aim for groups of the same size but with more candidates. The Better-matching Effect is illustrated in Figure 5.2, where users 1 and 2 are first grouped together; when new passengers emerge, they are separated and matched with other users such that the resulting routes get more efficient.

This effect is similar to the increase in "directness" in public transport systems reported by Fielbaum et al. (2020a), who argue that an increased number of passengers permits defining lines that require fewer detours because more passengers share the origins and destinations; however, there are relevant differences with ODRP as vehicles here do not follow fixed routes but adapt them online.

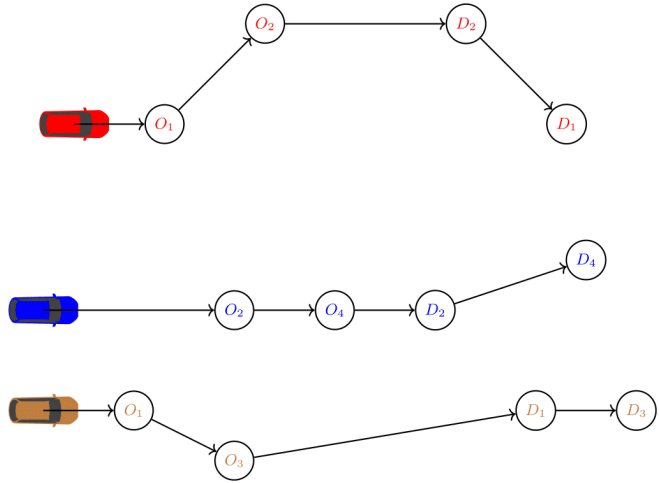

**Figure 5.2** Example of the Better-matching Effect. Both in the top row (low demand) and in the bottom row (high demand) we exhibit groups of size two. In the top row, the red vehicle is instructed to serve passengers 1 and 2, which are also marked with a red color. When the demand grows (bottom row), new passengers 3 and 4 appear, allowing the system to form more efficient groups. User 1 is now grouped with user 3 and served with a brown vehicle. Users 2 and 4 are grouped together to be served by a blue vehicle. The color of the passengers marks which vehicle serves them. Total delay decreases for the two users that remain from the top row, improving their perceived quality of service.

The presence of user-related scale phenomena is described in Figure 5.3, where we show the evolution of all the three sources discussed above as the demand grows. It is a stylized schematic figure that divides the analysis into three sectors, representing the respective zones (as seen in Figures 4.1 and 4.7a) in which the average delay first decreases, then increases, and then steadily decreases again. Figure 5.3 shows the so-called *Degree of scale economies*, which is formally defined for any production function as the average costs divided by the marginal costs: this means that there is a threshold in DSE = 1 determining whether scale economies or diseconomies prevails. The mentioned three sectors are:

- When the number of passengers is low (first sector of the curve), users hardly share a vehicle, so that the Flex-route and the Better-matching Effects are almost non-existent. This means that the Mohring Effect (which is more prominent when the demand is low) prevails, and there are economies of scale.
- Eventually, users begin to share the vehicle, and the system enters into the second zone. The Flex-route Effect begins to operate, and diseconomies of scale prevail. The Mohring Effect is still present. The Better-matching Effect also starts to operate but mildly due to the increased load (this effect operates directly when the number of users per vehicle is constant). Therefore, they are outweighed by the Flex-route Effect. The minimum of the curve represents the point at which vehicles' load increases at the fastest pace.
- Finally, when the vehicles cannot carry more passengers (they are full), the Flex-route Effect disappears, and the Mohring Effect has little impact. The Better-matching Effect, on the other hand, is fully operative, leading to DSE > 1. Eventually, DSE converges to 1 as all these sources get exhausted.

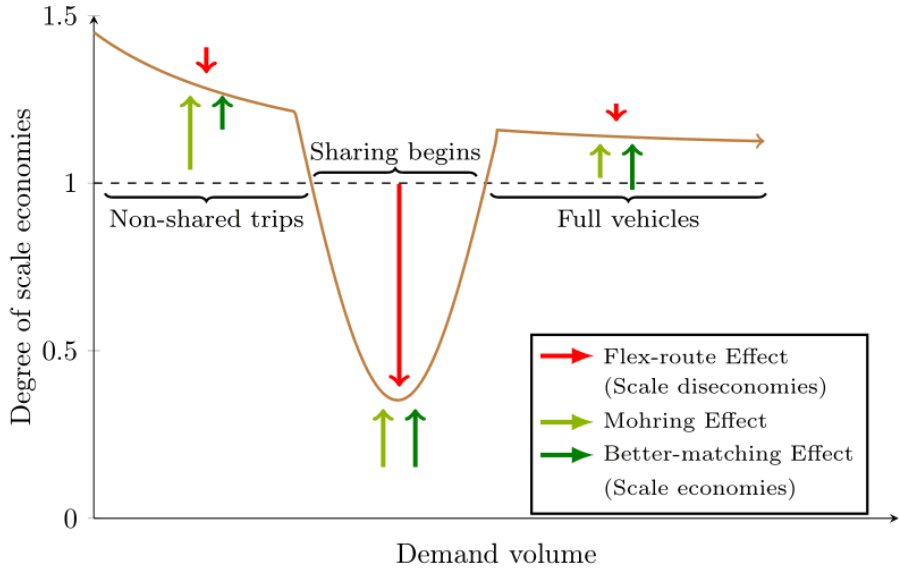

**Figure 5.3** Synthesis of the three sources of users-related scale effects for ODRP systems. The y-axis represents the degree of scale economies (DSE), so that scale economies prevail when DSE > 1 and the contrary happens when DSE < 1 (constant returns to scale if DSE = 1); the x-axis represents the number of users, and we do not include concrete numbers because this is a schematic representation. The direction of each arrow represents if it pushes DSE upwards (i.e., scale economies) or downwards (i.e., scale diseconomies), while its length represents its magnitude.

Such scale effects can be utilized to write mathematical relationships that define the users' cost function in ODRP. Their costs depend on the number of passengers $Y$, the average load of the vehicles $\varrho$, and the fleet size $B$. The system can decide the last two, and if we denote by $B(Y)$, $\varrho(Y)$ the optimal value for these variables as a function of $Y$, it naturally happens that $B' > 0$, $\varrho' > 0$ (note that as $\varrho$ increases, larger vehicles are required). Users' costs are defined by average waiting $\overline{t_w}$, walking $\overline{t_a}$, and in-vehicle time $\overline{t_v}$, all of them depending on $Y$, $B$, and $\varrho$. The Mohring Effect states that:

$$\frac{\partial \overline{t_h}}{\partial B} \leq 0 \text{ for } h = w, a \tag{8}$$

The Flex-route Effect implies that:

$$\frac{\partial \overline{t_h}}{\partial \varrho} \geq 0 \text{ for } h = w, a, v \tag{9}$$

And the Better-matching Effect states that:

$$\frac{\partial \overline{t_h}}{\partial Y} \leq 0 \text{ for } h = w, a, v \tag{10}$$

As $B$ and $\varrho$ increase with $Y$, the combined effect of all the three sources of scale can be positive or negative (due to the chain rule). Figure 5.3 synthesizes when each of these effects prevails, mainly depending on the varying rate at which $\varrho$ grows.

Regarding the comparison with public transport, our findings suggest that ODRP systems can play a better role when the demand is low because otherwise, the required fleet becomes too large due to the small capacity of the vehicles. However, the presence of scale economies suggests that for larger demand levels, they might also improve the public transport system, but not by replacing the whole line.

## 6. Conclusions and future research

In this paper, we have used the single-line model to understand the sources of scale economies and diseconomies in on-demand ridepooling (ODRP) systems that operate in the equivalent of a zone covered by a public transport line. To do this, we have extended a state-of-the-art assignment method for ODRP, in order to optimize the fleet size together with the decisions of how to group the users and which vehicle carries them.

Our simulations revealed that ODRP users induce both positive and negative externalities to the other passengers. Positive externalities are the Mohring Effect and the "Better-matching Effect", i.e., that it is possible to form more efficient groups when the demand grows; the negative externalities relate with increasing the number of users per vehicle, which induces longer detours, which we call the "Flex-routes Effect". There are only positive externalities on the operators' side, namely that vehicles can be used more intensely so that the fleet size grows less than linearly.

We have found that for the efficient operation of ODRP in a setting without request rejections, the possibility of asking the passengers to perform short walks to pick up points is crucial to keep total costs down, both for users and operators. In particular, we have found that an ODRP system with human-driven vehicles and walks allowed has a total cost at a similar level to that of a door-to-door ODRP system with automated (fully driverless case) vehicles. This finding has significant implications for the current and future design of mobility systems based on shared vehicles and shared rides, either with human-driven or automated vehicles.

If the system designer has to choose between a traditional transit line or an ODRP system, the latter should be mostly preferred for low-demand zones. However, the scale effects in ODRP suggest that there could be other ways of integrating both systems to enhance public transport and attract users from private modes in high-demand scenarios, especially for feeder-like systems. Understanding how this could be done is the most relevant future research question that emerges from this paper. Moreover, introducing the spatial components, i.e., looking at the whole transit network rather than at a single line, might reveal other sources of scale that do not show up here. Finally, considering for-profit companies, and how they compete, might have an influence on scale analysis that is also worth studying.

## Acknowledgments

Alejandro Tirachini acknowledges support from ANID Chile (Grant PIA/BASAL AFB180003).

# Appendix

## A.1 Public transport model

In order to compare the performance of the ODRP and the public transport systems, we now describe the public transport model we assume, following the classical model by Jansson (1980) and the posterior adaptations by Jara-Díaz & Gschwender (2009). We will describe in detail the circular model only, as the feeder one can be derived directly. Let us begin introducing some notation: $T$ refers to the time required by a bus to tour the whole circuit, i.e.

$$T = \frac{Z \cdot a \cdot L}{v_1} \tag{A1}$$

Where $L$ stands for the length of each arc. We assume that each user requires an average time $t$ to board and alight the bus. Denoting by $f$ the line frequency (to be optimized) and by $Y$ the number of passengers per time unit, then the bus cycle time is:

$$t_c = T + \frac{tY}{f} \tag{A2}$$

To use Eq. (A2) to express the operators' costs, we again follow Tirachini & Hensher (2011) and Jara-Díaz et al. (2017) to express both $c_O$ and $c_A$ (now read as operating costs per time-unit and capital costs, respectively) as values that grow linearly with the vehicles' capacity $K$, i.e.

$$c_O = c_{O1} + c_{O2}K, \quad c_A = c_{A1} + c_{A2}K \tag{A3}$$

As the operating time is fixed in the public transport case (buses are operating all the time), Eq. (A3) means that each bus cost can be expressed as $c_1 + c_2K$ (mind that this linear growth with respect to $K$ is also assumed for ODRP), with $c_1 = c_{A1} + Ec_{O1}$, $c_2 = c_{A2} + Ec_{O2}$, where $E$ is the total operation time. Operators' costs can then be written as:

$$f(T + \tfrac{tY}{f})(c_1 + c_2 K) \tag{A4}$$

Users' costs encompass waiting, in-vehicle, and walking times, as in Eq. (A5). They are valued differently by the users, with the respective parameters $p_w$, $p_v$, and $p_A$. Therefore, the public transport costs are calculated by solving the following optimization problem:

$$\min_{f,K} f(T + \tfrac{tY}{f})(c_1 + c_2 K) + Y(p_w \overline{t_w} + p_v \overline{t_v} + p_a \overline{t_a}) \tag{A5}$$

$$\text{s.t. } K \geq \tfrac{Y}{f}\alpha \tag{A6}$$

Eq. (A5) represents the sum of operators' and users' costs. We assume homogeneous headway, vehicles do not run full (passengers can board the first vehicle that arrives) and random user arrivals at constant rates, which imply that the average waiting time is $\overline{t_w} = f/2$. Average in-vehicle time $\overline{t_v}$ can be calculated as we know the average distance traveled by the users; it includes running time plus time spent at stops where other users board and alight. Average walking distance can be computed directly when the random demand is created, by calculating the distances between the real origins and the bus stations of the respective zones, and doing the same for the destinations. Dividing such distances by the walking speed $v_a$ results in the average walking time $\overline{t_a}$. Eq. (A6) ensures that all users will fit on the bus. As the objective function in Eq. (A5) increases with $K$, this constraint will always be active. Factor $\alpha$ represents the ratio between the most loaded and the average arc, which can also be computed directly once the random demand is known.

## A.2 Glossary and numerical value of the parameters

| Symbol | Meaning | Value |
| --- | --- | --- |
| $\delta$ | Time elapsed between two consecutive assignments in ODRP. | 1 [min] |
| $\tau$ | Time spent by the ODRP vehicle at each stop. | 10.5 [sec] |
| $a$ | Number of longitudinal streets in a zone. | 5 |
| $b$ | Number of transversal streets in a zone. | 7 |
| $v_1$ | Vehicles' speed in fast streets. | 25 [km/h] |
| $v_2$ | Vehicles' speed in low streets. | 12.5 [km/h] |
| $Z$ | Number of zones | 45 |
| $\gamma$ | Level of dispersion of the origins and destinations within a zone. | 0.2 |
| $l$ | Average number of zones toured by the users in the circular model. | 10 |
| $\sigma^2$ | Variance of the number of zones toured by the users in the circular model. | 4 |
| $L$ | Arcs' length. | 50 [m] |
| $t$ | Time required to board and alight a public transport vehicle. | 5 [sec] |
| $E$ | Total operation time | 10 [h] |
| $c_{O1}$ | Fixed operating cost per vehicle. | 1.13 [US$/min] |
| $c_{O2}$ | Capacity-dependant operating cost per vehicle. | 0.074 [US$/min-seat] |
| $c_{A1}$ | Fixed capital cost per vehicle (AV/Human-Driven). | 24.6/78.1 [US$] |
| $c_{A2}$ | Capacity-dependant capital cost per vehicle (AV/Human-Driven). | 2.1/1.2 [US$/seat] |
| $v_a$ | Walking speed. | 5 [km/h] |
| $p_v$ | Monetary equivalent cost of one time unit spent by a user in-vehicle. | 2.32 [US$/h] |
| $p_w$ | Monetary equivalent cost of one time unit spent by a user waiting. | 4.64 [US$/h] |
| $p_a$ | Monetary equivalent cost of one time unit spent by a user walking. | 4.64 [US$/h] |

**Table A1:** Glossary of the parameters used throughout the paper. Stopping time $\tau$ is computed following Roess et al. (2004). Operators' cost parameters $c_{O1}, c_{O2}, c_{A1}, c_{A2}$ for human-driven and automated vehicles are calculated for Santiago, Chile, based on Tirachini & Antoniou (2020). Time required to board and alight the vehicles $t$ is taken from Jara-Díaz et al. (2017). Walking speed $v_a$, as well as users' costs parameters $p_w$, $p_v$ and $p_a$ are obtained from Fielbaum et al. (2021). The rest of the parameters are ours.